\documentclass[twocolumn,showpacs,pre,superscriptaddress,floatfix]{revtex4-1}
\usepackage{amssymb,amsmath,amsfonts} \usepackage{bm} 
\usepackage{graphicx, color} 

\definecolor{Myorange}{cmyk}{0,0.42,1,0}

\newcommand{\avg}[1]{\langle #1 \rangle}

\begin{document}
\title{Non-parametric resampling of random walks for spectral network clustering}
\author{Fabrizio De Vico Fallani}
\affiliation{CNRS UMR-7225, H\^{o}pital de la Piti\'e-Salp\^{e}tri\`{e}re. Paris, France}

\author{Vincenzo Nicosia}
\affiliation{School of Mathematical Sciences, Queen Mary University of
  London, Mile End Road, E1 4NS, London (UK).}

\author{Vito Latora}
\affiliation{School of Mathematical Sciences, Queen Mary University of
  London, Mile End Road, E1 4NS, London (UK).}
\affiliation{Dipartimento di Fisica e Astronomia, Universit\'a di
  Catania, Via S. Sofia 61, 95123, Catania (Italy)}

\author{Mario Chavez}
\affiliation{CNRS UMR-7225, H\^{o}pital de la Piti\'e-Salp\^{e}tri\`{e}re. Paris, France}

\date{\today}
\begin{abstract}
Parametric resampling schemes have been recently introduced in complex
network analysis with the aim of assessing the statistical
significance of graph clustering and the robustness of community
partitions.  We propose here a method to replicate structural features
of complex networks based on the non-parametric resampling of the
transition matrix associated with an unbiased random walk on the
graph. We test this bootstrapping technique on synthetic and
real-world modular networks and we show that the ensemble of
replicates obtained through resampling can be used to improve the
performance of standard spectral algorithms for community detection.
\end{abstract} 

\pacs{89.75.-k,  02.50.Ga, 05.10.Ln}
\maketitle %

In the past decade, network science has proven to be a robust and
comprehensive framework to investigate, model and understand the
structure and function of the complex interaction patterns observed in
diverse biological, physical, social and technological
systems~\cite{boccaletti,Barabasi2002rev,Barrat2008,Newman2010}. One
of the most intriguing characteristic of many real-world complex
networks is the presence of communities, i.e. tightly-knit groups of
nodes which exhibit poor connectivity with the rest of the
graph~\cite{Girvan2002}. As a matter of fact, many experimental
evidences have confirmed that communities are the meso-scale building
blocks of complex networks: they usually correspond to functional
modules in the brain~\cite{Bullmore2009,chavez}, to topical clusters
in social and communication networks~\cite{Newman2004}, to metabolic
reactions and functional domains in protein interaction
networks~\cite{Guimera2005,Jonsson2006}, to disciplines and research
areas in collaboration networks~\cite{Girvan2002}. Consequently, a lot
of effort has been devoted to the identification of efficient
algorithms for community detection~\cite{fortunato}.

A typical problem in complex networks analysis is that a real-world
network is just a single observation drawn from an unknown
distribution of graphs having certain
characteristics~\cite{wassermanBOOK}. As a consequence, there is no
predefined way to assess the statistical variability of any local,
meso-scale of global network property, including the presence and
composition of communities.
A widely used approach to determine the statistical significance of
network observables consists in considering random \textit{network
  ensembles}, i.e. sets of graphs obtained from the original network
by keeping fixed some structural properties (e.g. the degree sequence
or the clustering coefficient) and rewiring the edges at
random~\cite{maslov,ziv,bianconi2}. In the case of community
detection, this approach led to the definition of the modularity
function, which quantifies the significance of a given community
partition of a graph as the deviation from the average modularity
expected in an ensemble of random graphs having the same degree
sequence~\cite{Newman2004}.
Another possibility is \emph{parametric bootstrapping}, in which the
robusteness of a network property is assessed against small
perturbations of the graph connectivity~\cite{gfeller,karrer,rosvall}.
This approach relies on the hypothesis that the observed network is
representative of a set of graphs (a model) having a certain
(\textit{a-priori} known) structure. Consequently, the variability of
any network observable can be estimated as the deviation from the
average of the corresponding model. Many different parametric
resampling schemes have been used to assess the robustness of network
communities against small connectivity perturbations. However, all
these methods require an a-priori hypothesis about the model to which
the network belongs, so that the unbiased statistical assessment of a
network partition remains an open challenge~\cite{mirshahvalad}.

A possible solution to this problem is \textit{non-parametric
  bootstrapping}, a data-driven technique for providing the
statistical confidence of almost any statistical
estimate~\cite{efronBOOK,efron}, based on the generation of repeated
observations (replicates) from an unknown population using the
available data samples (in our case, a single network) as a starting
point. This approach has been successfully employed for several
different applications, and in particular to improve the stability and
accuracy of clustering algorithms in metric spaces~\cite{Strehl02}.

In this Brief Report we propose to use non-parametric bootstrapping to
improve the performance of spectral community detection
algorithms. The method is based on the construction of replicates of
the transition matrix of the network, and on the estimation of an
average distance matrix, whose elements correspond to the expected
spectral distances between pairs of nodes of the graph, averaged over
the ensemble of replicates. Then, the obtained distance matrix is fed
into a standard hierarchical clustering algorithm. The idea is that
the aggregation of information about different replicates,
representative of the unknown ensemble to which the network belongs,
should filter out noise and allow to obtain more accurate and robust
partitions than the one found on the original network.  This approach
is in the same line of ensemble or consensus clustering methods, which
combine several partitions generated by different clustering
algorithms ---or by different runs of the same algorithm--- into a
single, more robust
partition~\cite{Kwak2009,karrer,rosvall,kim,lancichinetti12}. We
analyze the community partitions obtained by non-parametric
bootstrapping in different synthetic and real-world modular networks,
and we show that this approach can substantially improve the
performances of existing spectral clustering methods.


\emph{Spectral clustering for community detection.--} Let $G(V, E)$ be
a connected undirected and unweighted graph with $N=|V|$ nodes and
$K=|E|$ edges, and let $A=\{a_{ij}\}$ be the adjacency matrix of $G$,
whose entry $a_{ij}=1$ if there is an edge connecting node $i$ and
node $j$, while $a_{ij}=0$ otherwise. We consider the problem of
finding communities of $G$, i.e. subsets of nodes of $G$ which are
more connected internally than with the rest of the
graph~\cite{Newman2004}. Several community detection algorithms are
based on mapping each node of $G$ into a point of an appropriate
metric space $X$, so that two nodes $i$ and $j$ having similar
structural properties (e.g., similar set of neighbours) are mapped to
two points $x_i$ and $x_j$ placed relatively close to each other in
$X$. Then, the nodes are clustered according to the Euclidean distance
between their corresponding images in $X$, so that nodes whose
projections are closer in $X$ have a higher probability to be put in
the same cluster.

A widely adopted method to map the nodes of a graph into a metric
space makes use of spectral properties of matrices associated to $G$,
and in particular of the eigenvectors of the adjacency matrix $A$ or,
more frequently, of the transition matrix $P=\{P_{ij}\}$ associated to
an unbiased random walk on the graph ($P_{ij}=a_{ij}/k_i$, where
$k_i=\sum_j a_{ij}$ is the degree of node
$i$)~\cite{gfeller2,chavez,cvetkovicBook,chungBook}. This choice is
motivated by the observation that both $A$ and $P$ carry information
about the overall structure of the graph.
Here we consider the transition matrix $P$. This matrix is
characterized by a set of eigenvalues $\{\lambda_0,
\lambda_1,\ldots,\lambda_{N-1}\}$ such that $|\lambda_0| \geq
|\lambda_1| \geq \ldots\geq |\lambda_{N-1}|$. Each eigenvalue
$\lambda_k$ is associated to the left and right eigenvectors
$\varphi_k$ and $\psi_k$, which satisfy $\varphi_k^{\intercal}{P} =
\lambda_k \varphi_k^{\intercal}$ and ${P}\psi_k=\lambda_k \psi_k$,
respectively.
Thus, it is possible to map node $i$ to the point $x_i\in
\mathbb{R}^{N}$ whose $k^{\text {th}}$ coordinate is equal to the the
$i^{\text {th}}$ component of the $k^{\text {th}}$ right eigenvector
of ${P}$.  The distance $d_{ij}$ between two points $x_i$ and $x_j$,
can be written in terms of eigenvectors and eigenvalues of
$P$~\cite{lafon2006}, namely:
%
$  d_{ij} = \sqrt{\sum_{k \geq 1}\lambda_k^{2}(\psi_k(i) - \psi_k(j))^2}$
%
where $\psi_k(j)$ denotes the $j^{\text{th}}$ component of the
$k^{\text{th}}$ right eigenvector. In general, this distance can be
approximated by using only the first $\beta$ nontrivial eigenvectors
and eigenvalues of $P$~\cite{note_approx}:
\begin{equation}
  d_{ij} \simeq \sqrt{\sum_{k = 1}^{\beta} \lambda_k^{2}(\psi_k(i) -
  \psi_k(j))^2}.
  \label{eq:embedding}
\end{equation}
The elements $\{d_{ij}\}$ of the matrix ${D}$ obtained
from Eq.~(\ref{eq:embedding}) represent the distances between each
pair of points $x_i$ and $x_j$ in the lower dimensional space
$X\equiv\mathbb{R}^ \beta $.

\begin{figure*}[!ht]
   \centering
   \includegraphics[width=6in]{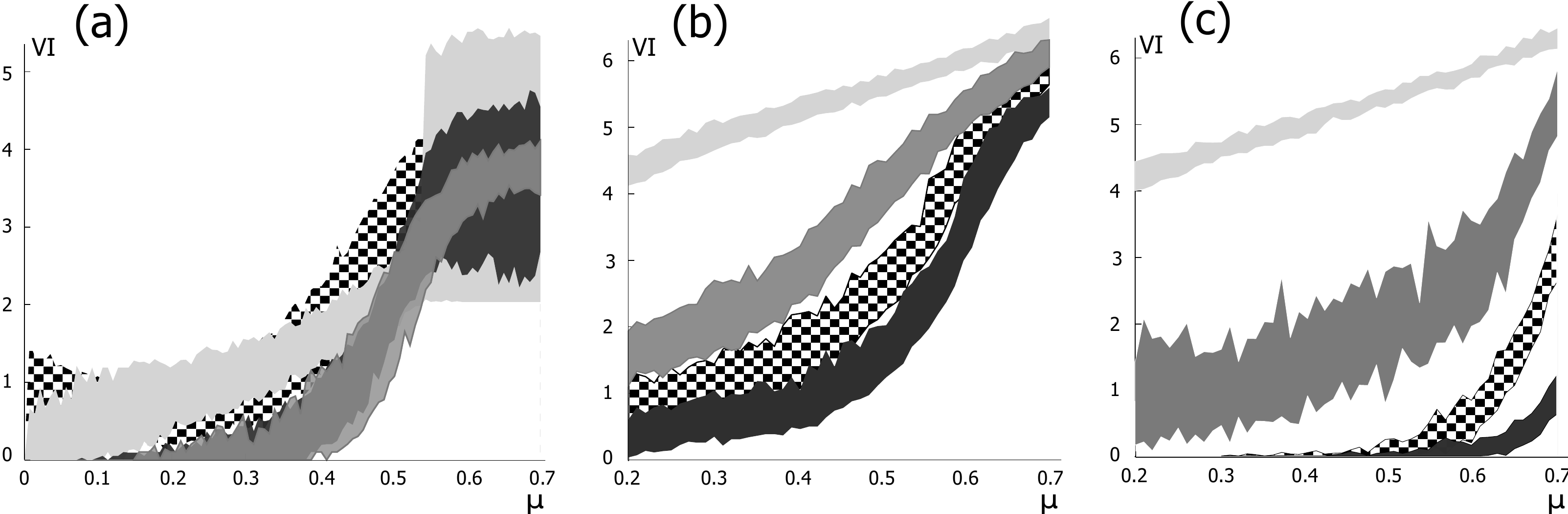}
   \caption{\textbf{Benchmark networks.} The variation of information
     $VI$ as a function of the proportion of inter-modules links $\mu$
     in GN graphs (a) and as a function of the mixing parameter $\mu$
     in LFR500 (b) and LFR2000 (c) graphs. The region inside each
     curve includes the $5^{th}$ and the $95^{th}$ percentiles of $VI$
     across $R$ different runs. The four curves in each panel
     correspond to the optimal partitions obtained using,
     respectively, the distance matrix ${D}$ induced by ${P}$ for
     $\beta=1$ (light gray) and $\beta=10$ (checked pattern), the
     average distance matrix ${\widetilde{D}}^{*}$ over the population
     of replicates ${P}^*$ for $\beta=1$ (black), and modularity
     optimization on the adjacency matrix $A$, as described in
     Refs.~\cite{newmanEigenvectorsPNAS,Newman2004} (dark-gray). The
     network order $N$, the number of runs $R$, and the number $B$ of
     bootstrap realizations for each run are (a) $N=128$, $R=100$, and
     $B=100$; (b) $N=500$, $R=100$, and $B=100$; and (c) $N=2000$,
     $R=50$, $B=50$.}
   \label{fig:fig1}
\end{figure*}

Since the terms $\{d_{ij}\}$ represent distances in a metric space,
then we can use the matrix $D$ to detect candidate community
partitions of $G$ by means of hierarchical clustering, an iterative
aggregation algorithm which starts by considering each node as a
separate cluster, and successively merges the two clusters having
minimal distance according to $D$~\cite{note_linkage}. The algorithm
stops when all the nodes have been grouped in a single cluster.
The hierarchical clustering algorithm produces a dendrogram $H$,
i.e. a tree where each of the $N-1$ internal nodes represents the
fusion of two clusters. A horizontal cut of $H$ corresponds to a
partition of the graph into a certain number of communities. The
quality of each partition $\mathcal{S}$ can be quantified using the
modularity function~\cite{Newman2004}, which compares the abundance of
edges lying inside each community with respect to a null model. In
formula:
\begin{equation}
  \mathcal{Q}(\mathcal{S})=\sum_{s=1}^{N_s}{\left[{\frac{m_{s}}{K}-\left(\frac{k_{s}}{2K}\right)}^{2}\right]},
\end{equation}
where $N_s$ is the number of clusters in the partition $\mathcal{S}$,
$K$ is the total number of edges in the network, $m_{s}$ is the number
of edges between vertices in cluster $s$, and $k_{s}$ is the sum of
the degrees of the nodes in cluster $s$. We assume that the best
partition in communities of the graph $G$ is the cut of the dendrogram
$H$ having maximum modularity.

\textit{Clustering through non-parametric bootstrapping.-- } The main
limitation of community detection algorithms based on the spectral
properties of the transition matrix is that the obtained partition is
pretty sensitive to fluctuations in the adjacency matrix of the
graph. As a matter of fact, the eigenvalues and eigenvectors of the
transition matrix can be substantially modified by adding, deleting or
rewiring just a few edges. Therefore, we propose to improve the
quality of spectral clustering by using information about the average
spectral properties of transition matrices obtained by a
non-parametric bootstrapping of the observed matrix $P$.

The authors of Ref.~\cite{basawa} have proposed a generic bootstrap
scheme to resample the transition probabilities of a finite state
time-invariant Markov chain. Starting from a realization $\chi$ of the
Markov chain, one constructs the maximum likelihood estimator of the
associated transition matrix $P$ as $p_{ij}=\frac{f_{ij}}{f_i}$, where
$f_{ij}$ is the observed number of transitions from state $i$ to state
$j$ in $\chi$ and $f_i=\sum_j f_{ij}$.  Then, replicates of the
observed transition matrix are obtained by drawing, for each state
$i$, the random variables $\{f^*_{i1}, \ldots, f^*_{iN}\}\sim
\text{Multinomial}(f_i; p_{i1}, \ldots,p_{iN})$ according to
$\widetilde{P_{ij}}=\frac{f^*_{ij}}{f_i}$. The distribution of
${\widetilde{P}}$ is then obtained by Monte-Carlo sampling. This
approach was shown to be asymptotically valid for approximating the
sampling distribution of ${P}$~\cite{basawa}, and has been also used
to assess the confidence intervals of transition probabilities in
disease modeling~\cite{sendi}.

Since the unbiased random walk on the graph $G$ defined by the
transition matrix $P$ is a finite-state time-invariant Markov chain,
we can construct a similar resampling scheme in which replicates of
the transition matrix ${P}$ are obtained by randomly drawing the
variables $\{f^*_{i1}, \ldots, f^*_{iN}\}$ from a multinomial
distribution with probabilities $\{p_{i1}, \ldots,p_{iN}\}$,
conditional on the observed degree sequence $\{k_i\}$ of $G$. It is
worth noticing that, in contrast to previous approaches where each
link was resampled independently from the
others~\cite{gfeller,rosvall}, here the replicas of the transition
probabilities for each node $i$ are drawn from a multinomial
distribution, accounting for the observed transitions to other nodes
$\{p_{i1}, \ldots,p_{iN}\}$.

Given the transition matrix ${P}$ of $G$, we generate $B$ bootstrap
transition matrices $\{{P^*_1}, {P^*_2}, \ldots, {P^*_B}\}$. Then, we
project each matrix ${P^*_b}$ into $\mathbb{R}^{\beta}$ (where $\beta$
is a tunable parameter), and we estimate the corresponding bootstrap
distance matrices ${D^*_b}$, whose entry $d^{b}_{ij}$ is the Euclidean
distance between $x_i$ and $x_j$ in $\mathbb{R}^{\beta}$ according to
the mapping induced by $P^{*}_{b}$. Then, we compute the average
distance matrix ${\widetilde{D}^*}=\frac{1}{B} \sum_b {D^*_b}$, which
is expected to be the most consistent (similar) with the central
tendency of the population of replicates.
The matrix ${\widetilde{D}^*}=\{\widetilde{d}^*_{i,j}\}$ effectively
quantifies the dissimilarity between any pair of vertices of $G$ (the
smaller the distance $\widetilde{d}^*_{i,j}$ the more similar are $i$
and $j$), in terms of the average distance between their projections
in $\mathbb{R}^{\beta}$ across several replicas of ${P}$.

We notice that, in principle, the spectral network decomposition based
on non-parametric bootstrapping does not rely on modularity, so that
any quality function can be used to determine the best partition in
the dendrogram. We also would like to stress that the partitions
obtained with resampling-based clustering methods do not necessarily
provide the absolute optimum of a given quality function. Instead,
non-parametric bootstrapping yields partitions that are the most
consistent with the central tendency across different replicates drawn
from the same population.

\emph{Synthetic networks.--} We have tested the performance of our
approach on two classes of synthetic graphs with tunable modular
structure. In the first benchmark (GN), proposed by Girvan and
Newman~\cite{Girvan2002}, each network consists of $N=128$ nodes
divided into $4$ modules of equal size. Pairs of nodes in the same
module are connected with probability $p_{in}$, while nodes belonging
to different modules are linked with a probability
$p_{out}$. Parameters are set such that the average degree is kept
constant $\avg{k}=16$. By appropriately tuning $p_{in}$ and $p_{out}$
one can set the percentage $\mu$ of edges lying between
communities. The second class of modular graphs (LFR), proposed by
Lancichinetti, Fortunato and Radicchi~\cite{lancichinetti}, accounts
for the heterogeneity in the distributions of node degrees and
community sizes. In this case, we generated modular graphs with
scale-free degree distribution $P(k)\sim k^{-\gamma}$ and community
size distribution $P(s) \sim s^{-\eta}$, where $\gamma=2$ and
$\eta=1$. An appropriate tuning of the model parameters allows to
create graphs with a prescribed fraction $\mu$ of inter-community
edges. We considered graphs having $N=500$ and $\avg{k}=7$ (LFR500)
and graphs with $N=2000$ nodes and $\avg{k}=28$ (LFR2000). 

Since the real partition in communities of these synthetic networks is
a-priori known, we can compare the best partition obtained through
spectral clustering with the reference one. A widely used measure to
compare two different partitions is the variation of information
($VI$)~\cite{meila}. In a nutshell, this non-negative metric
quantifies how much information is lost and gained in changing from a
partition $\mathcal{A}$ to a partition $\mathcal{B}$. It can be
estimated by $VI(\mathcal{A},\mathcal{B}) =
-\sum_i^{c^\mathcal{A}}\sum_j^{c^\mathcal{B}}(\frac{n^{\mathcal{AB}}_{ij}}{N})\log
\frac{n^{\mathcal{AB}}_{ij}}{N} + \frac{n^{\mathcal{AB}}_{ij}}{N}\log
\frac{{\mathcal{AB}}_{ij}/N}{n^\mathcal{A}_i n^\mathcal{B}_j /N^2}$,
where $c^\mathcal{A}$ ($c^\mathcal{B}$) is the total number of
clusters in the partition $\mathcal{A}$ ($\mathcal{B}$),
$n^\mathcal{A}_i$ ($n^\mathcal{B}_j$) is the number of nodes in the
$i^{\text{th}}$ ($j^{\text{th}}$) cluster of partition $\mathcal{A}$
($\mathcal{B}$), and $n^{\mathcal{AB}}_{ij}$ is the number of nodes
shared by the $i^{\text{th}}$ cluster of partition $\mathcal{A}$ and
the $j^{\text{th}}$ cluster of partition $\mathcal{B}$. Values of $VI$
range from $0$, when $\mathcal{A}$ and $\mathcal{B}$ are identical
partitions, to $\log N$ when both $\mathcal{A}$ and $\mathcal{B}$ are
randomly drawn.
\begin{figure}[!t]
   \centering
   \includegraphics[width=3in]{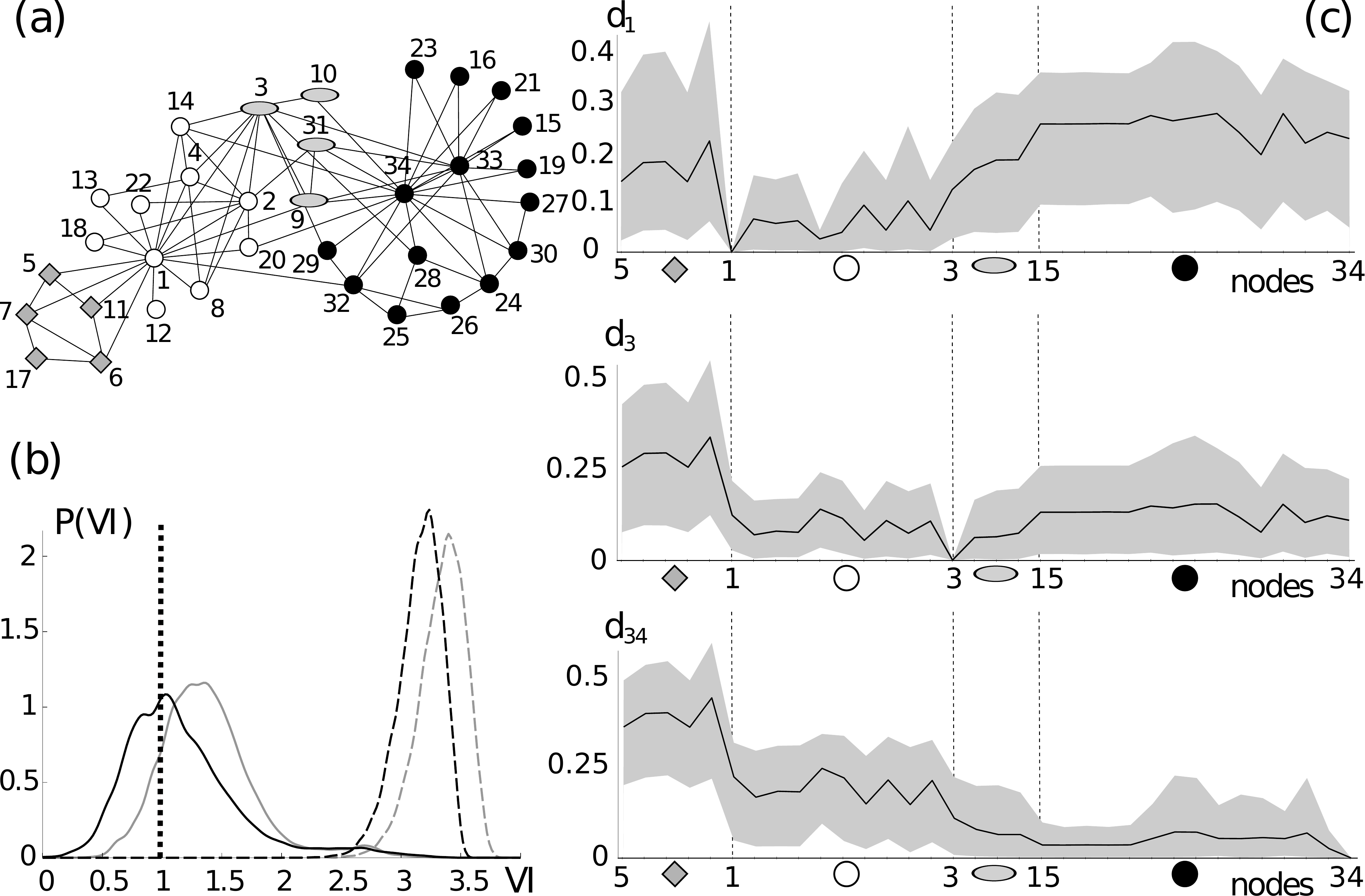}
   \caption{ (a) The best partition of the Zachary Karate club network
     obtained through non-parametric bootstrap ($B=20000$ replicas)
     gives a value of modularity $Q=0.389$; (b) the distribution of
     $VI$ across the replicates with respect to the partition induced
     by $\widetilde{D}^*$ (solid black line) and the partition with
     maximum modularity (solid grey line); the vertical dotted line
     indicates the $VI$ between the partition with maximum modularity
     and the one induced by $\widetilde{D}^{*}$ ($VI=0.952)$. Dashed
     lines indicate the distribution of $VI$ for $B=20000$ random
     partitions, with respect to the partition induced by
     $\widetilde{D}^*$ (dashed black line) and the one with maximum
     modularity (dashed grey line). (c) Spectral distance between node
     $i=1$, (top) $i=3$ (middle), $i=34$ (bottom) and the rest of the
     nodes. Gray regions indicate the
     $0.05^{\text{th}}$--$95^{\text{th}}$ percentiles interval of the
     bootstrap distribution.}
   \label{fig:fig2}
 \end{figure}
%

Fig.~\ref{fig:fig1} shows the variation of information between the
reference partition and the best one obtained through bootstrap-based
spectral clustering, as a function of the fraction of inter-community
edges $\mu$. The reported results suggest that even when the graphs do
not have any more a strong community structure, i.e. when for each
node the number of neighbours outside its community is similar with
the number of neighbours inside its community, the accuracy of the
proposed bootstrap-based method remains pretty high.  For GN networks,
the accuracy of the bootstrap-based method is comparable to that of a
standard modularity optimization
algorithm~\cite{newmanEigenvectorsPNAS,Newman2004}. For LFR500 and
LFR2000, the non-parametric bootstrap method outperforms the other
algorithms, even when we consider an embedding with $\beta=1$, and
exhibits a smaller value of $VI$ up to relatively large values of
$\mu$.

\emph{The Zachary's karate club network.--} Fig.~\ref{fig:fig2}(a)
shows the best partition found by the bootstrap-based algorithm
($B=20000$, $\beta=1$) in the Zachary's karate club
network~\cite{zachary}, a paradigmatic example of graph with a strong
modular structure. The partition consists of three main modules (black
circles, white circles and grey diamonds, respectively) and a small
interface community which contains nodes $\{3,9,10,31\}$ (grey
ellipses). The distribution of the $VI$ between the partitions
obtained through spectral clustering on each single replicate and the
one found using ${\widetilde{D}^*}$ (reported in
Fig.~\ref{fig:fig2}(b), solid black line) shows that the latter one
indeed represents the central tendency of the population of
replicates. However, the typical $VI$ between the partition of a
single replicate and the one with maximum modularity
(Fig.~\ref{fig:fig2}(b), grey curve) is higher than that obtained by
averaging over all replicates (indicated in Fig.~\ref{fig:fig2}(b) by
the vertical line). Notice also that the typical $VI$ between random
partitions of the graph and, respectively, the one obtained averaging
over all replicates or the one with maximum modularity (respectively
the dashed black line and the dashed grey line in
Fig.~\ref{fig:fig2}(b)) is much larger than that obtained through
spectral clustering.

Despite the partition of Fig.~\ref{fig:fig2}(a) is not the one with
maximum modularity~\cite{duch2005}, it is worth noticing that most of
the nodes put in the interface community (namely, $3$, $9$ and $10$)
have been ambiguously classified by many different community detection
algorithms~\cite{li,estrada}, mostly because assigning them to either
the black or the white module has negligible effects on
modularity~\cite{gfeller}. A more in-depth analysis of the
dissimilarity matrix ${\widetilde{D}^*}$ provides a possible
explanation for this fact. In Fig.~\ref{fig:fig2}(c) we report the
average spectral distance $\widetilde{d}^*_{ij}$ between node $1$ (top
panel), node $3$ (middle panel) and node $34$ (bottom panel) and all
the other nodes in the graph. As expected, both node $1$ and node $34$
exhibit a sensibly smaller distance towards the other nodes in their
repective natural communities, which is consistent with the fact that
$1$ and $34$ are known to be the centers of these two
groups. Conversely, the distance between node $3$ and the nodes in the
white community is comparable to that between $3$ and the nodes in the
black community. This explains why the central tendency of the
population of replicates is to place node $3$ in a separate community,
together with other three nodes having a similar spectral distance
pattern.

\emph{Concluding remarks.--} In this work we have shown how the
generation of replicates of the transition matrix associated to a
graph allows to improve the performance of community detection
algorithms based on spectral clustering. In general, we believe that
non-parametric bootstrapping techniques, which do not require any
assumption about the ensemble of graphs to which a given network
belongs, might be successfully employed also to assess the
significance of the variability of nodes attributes defined by
different random walk parameters (e.g. hitting times or return times),
and for the statistical validation of other structural properties
defined on different flavors of random
walks~\cite{allefeld07,sinatra10}.

\bigskip
This work was supported by the EU-LASAGNE Project, Contract No.318132
(STREP). F. De Vico Fallani is financially supported by the French
program ``Investissements d'avenir" ANR-10-IAIHU-06.

\end{document}